\documentclass[conference]{IEEEtran}
\IEEEoverridecommandlockouts
\usepackage{cite}
\usepackage{amsmath,amssymb,amsfonts}
\usepackage{algorithmic}
\usepackage{algorithm}
\usepackage{graphicx}
\usepackage{multirow}
\usepackage{rotating}
\usepackage{textcomp}
\def\BibTeX{{\rm B\kern-.05em{\sc i\kern-.025em b}\kern-.08em
    T\kern-.1667em\lower.7ex\hbox{E}\kern-.125emX}}

\begin{document}

\title{A Novel Image Encryption Scheme Based on \\ Different Block Sizes for Grayscale and \\ Color Images}

\author{
\IEEEauthorblockN{Omar Reyad}
\IEEEauthorblockA{Sohag University, Egypt\\
Emails: ormak4@yahoo.com}
\and 
\IEEEauthorblockN{M. A. Mofaddel}
\IEEEauthorblockA{Sohag University, Egypt\\
mmofaddel@hotmail.com}
\and
\IEEEauthorblockN{W. M. Abd-Elhafiez}
\IEEEauthorblockA{Sohag University, Egypt\\
w\_a\_led@yahoo.com}
\and
\IEEEauthorblockN{Mohamed Fathy}
\IEEEauthorblockA{Sohag University, Egypt\\
mohamed\_fathy@yahoo.com}
}

\maketitle

\begin{abstract}
In this paper, two image encryption schemes are proposed for grayscale and color images. The two encryption schemes are based on dividing each image into blocks of different sizes. In the first scheme, the two dimension ($2$D) input image is divided into various blocks of size $N \times N$. Each block is transformed into a one dimensional ($1$D) array by using the Zigzag pattern. Then, the exclusive or (XOR) logical operation is used to encrypt each block with the analogous secret key. In the second scheme, after the transformation process, the first block of each image is encrypted by the corresponding secret key. Then, before the next block is encrypted, it is XORed with the first encrypted block to become the next input
to the encrypting routine and so on. This feedback mechanism depends on the cipher block chaining (CBC) mode of operation which considers the heart of some ciphers because it is highly nonlinear. In the case of color images, the color component is separated into blocks with the same size and different secret keys. The used secret key sequences are generated from elliptic curves (EC) over a \textit{binary} finite field $\mathbb{F}_{2^{m}}$. Finally, the experimental results are carried out and security analysis of the ciphered images are demonstrated that the two proposed schemes had a better performance in terms of security, sensitivity and robustness. 
\end{abstract}

\section{Introduction} \label{intro}
With the great progress in Internet technology and the maturation of digital image processing techniques, various applications of digital images are commonly widespread and are still continuously and rapidly increasing in the future. Although, a potential information security risk is still and always exist during the transmission operation of digital images over an opened (unsecured) networks. Naturally, the security of multimedia content such as plainimages attracts more and more attention from cryptographers points of view and leads to the importance of image encryption technology in many applications. 

Image encryption principles is different from that of text due to some intrinsic features of images such as bulk data capacity, high redundancy, strong correlation among adjacent pixels, and so forth. The traditional of various encryption algorithms that is based on number theory such as Data Encryption Standard (DES), Advanced Encryption Standard (AES) and the RSA cryptosystem, are found to have the weakness of low-level efficiency when the image is becoming bulky and large \cite{c1}, \cite{c2}. Consequently, these algorithms are not fully suitable for the encryption of such kind of large sized data, especially for a real-time communication scenarios.

In many situations, the color images are commonly used and frequently transmitted over the Internet and through wireless communication networks, because, they contain more abundant information than the grayscale images. Though some encryption algorithms for grayscale images can be easily extended and modulated to handle color images, it consumes more running time due to additional information required to represent the Red, Green and Blue (RGB) color image components. So, the need to develop a secure encryption algorithm for color images has attracted growing attentions in recent years \cite{c3},\cite{c4}.

In this paper, we proposed a new encryption scheme based on segment the image into blocks of size $N \times N$, in order to increase the security level of the cipherimages and the resistance against known-plaintext and chosen-plaintext attacks. 

The paper is organized as follows. In Section \ref{Sec1}, the preliminaries of EC and CBC mode are discussed. In Section \ref{Sec2}, given a background about the current image encryption schemes. In Section \ref{Sec3}, the description of the proposed schemes are presented. In Section \ref{Sec4}, presented the experimental results and discussion while conclusions are given in Section \ref{Sec5}.

\section{Preliminaries} \label{Sec1}
\subsection{Elliptic Curve over a Binary Finite Field} \label{Sec1a}
The field $\mathbb{F}_{2^{m}}$ is called a \textit{binary} finite field and it can be viewed as a vector space of dimension $m$ over the field $\mathbb{F}_{2}$ which consists of two binary elements $\{0,1\}$. A non-supersingular elliptic curve $E$ over a binary field $\mathbb{F}_{2^{m}}$ is defined by an equation taking the form

\begin{equation} \label{EQ1.1}
y^{2} + xy = x^{3} + a x^{2} + b 
\end{equation} 

\noindent where the parameters $a, b \in \mathbb{F}_{2^{m}}$ with $b \ne 0$. The set $E(\mathbb{F}_{2^{m}})$ consists of all the points $(x, y), x \in \mathbb{F}_{2^{m}}, y \in \mathbb{F}_{2^{m}}$, which satisfy the defining equation given in (\ref{EQ1.1}), together with a special point $O$ called the point at infinity. These set of points form an abelian group with respect to the arithmetic of elliptic curve addition rules over this abelian group \cite{c5}.

\subsection{The Chaos-Driven ECPRNG} \label{Sec1b}
The Chaos-Driven Elliptic Curve Pseudo-random Number Generator (C-D ECPRNG) is considered to be the EC-Linear Congruential Generator \cite{c6} driven by a chaotic map and is presented in \cite{c7} for the prime finite field $\mathbb{F}_{p}$. Such specific modification improves randomness of the sequence generated and increases it's periodicity. The C-D ECPRNG for a given seed point $G(x,y) \in E(\mathbb{F}_{2^{m}})$ as the secret key, is defined as the following sequences of points generated by EC-points addition operation:

\begin{equation} \label{EQ1.6}
\begin{split}
U_{i} & = [i(1+b_{i})]G\oplus U_{0} \\
      & = \left\{\begin{array}{l c c} {[i]G\oplus U_{0}} & \; \text{if} & {b_{i} =0} \\ {[2i]G\oplus U_{0}} & \; \text{if } \; & {b_{i} =1} \end{array}\right. \quad , i=1,2,... 
\end{split}
\end{equation}

\noindent where $U_{0}(x,y) \in E(\mathbb{F}_{2^{m}})$ is the "initial value" and $b_{i} $ is the random bits generated by a chaotic map $\Phi $ 

\begin{equation} \label{EQ1.7}
b_{i} =\left\{\begin{array}{l c c} {0 } & \; \text{if} \; & {{\Phi}^{i} (s)\in S_0} \\ {1} & \; \text{if} \; & {{\Phi}^{i} (s)\in S_1} \end{array}\right. \quad  ,i=1,2,...  
\end{equation}

\noindent where the state space $S = [0,1]$ is the interval and $S_0 = \left[ 0,0.5 \right] $, $S_1 = \left( 0.5, 1 \right]$ are two subsets of the interval equal to $0.5$ \cite{c8}.

\subsection{The CBC Mode of Operation} \label{Sec1c}
In cipher block chaining (CBC) mode, the encryption of a block not only depends on the key but also on the previous blocks \cite{c9}. The encryption process is context-dependent operation. This means that the same size blocks in different contexts are encrypted differently. The receiver can confirm that the cipher block has been changed because the decryption process of a manipulated cipher block does not work.

The CBC mode uses a fixed initialization vector (IV) which can be made public. Then, the plainimage is decomposed into blocks of length $n$. If Alice encrypts the sequence $b_l, ... , b_n,$ of plainimage blocks of length $n$ using the key $k$, then she sets

\begin{equation} \label{EQ1.13}
C_0 = IV, \quad C_j = E_k(C_{j-1} \oplus B_j), \quad {1 \le j \le n} 
\end{equation}

She obtains the cipher block

\begin{equation} \label{EQ1.14}
C = C_1 . . . C_n
\end{equation}

In general, the CBC mode of operation encrypts the same plainimage differently with different initialization vectors. Moreover, the encryption of a
plainimage block depends on the preceding plainimage blocks. Therefore, if the order of the cipherimage blocks is changed or if the cipherimage
blocks are replaced, then the decryption process becomes very hard or even impossible.

\section{Literature Review} \label{Sec2}
Pareek et al. \cite{c3} proposed a new approach for color image encryption based on chaotic logistic maps. An external secret key of $80$-bit length and two chaotic logistic maps are employed. Eight different operation-types are used to encrypt the image pixels. In \cite{c10}, Huang and Nien proposed a color image cryptosystem using multi-chaotic systems, which is composed of two shuffling stages parameterized by chaotically generated sequences. But, it is found that this method cannot resist known-plaintext attack and chosen-plaintext attack \cite{c11}. Liu and Wang \cite{c12} designed a stream-cipher algorithm based on one-time keys and robust piecewise linear chaotic maps in order to get high security and improve the dynamical degradation. The initial conditions were generated by the Message Digest hash function (MD5) of the mouse positions. In \cite{c13}, Patidar et al. have designed a fast loss-less symmetric color image cipher based on the widely used substitution-diffusion principle which utilized chaotic logistic maps. In \cite{c14} Rhouma et al. have proposed an approach for color image encryption based on one-way coupled-map lattices (OCML). An external secret key of $192$-bit length was used to generate the initial conditions and parameters of the OCML by making algebraic transformations to the secret key. Liu and Wang \cite{c15} applied a bit-level permutation and high-dimension piecewise linear chaotic map to encrypt color image. The chaotic Chen system was employed to confuse and diffuse the red, green and blue components simultaneously. In \cite{c16}, Ye has proposed an efficient image encryption scheme based on generalized Arnold map and generalized Bernoulli shift map. The proposed scheme can shuffle the plainimage efficiently in the permutation process.  In \cite{c17}, a new color image encryption algorithm was presented based on Logistic map, which is used to encrypt the red, green and blue components of color image at the same time and make the three components affect each other. 

It is found that most of the chaos-based cryptosystems for color images usually employ the low dimensional and single chaotic system which leads to some fundamental drawbacks such as insufficient key space and weak security function. Moreover, the red, green and blue components are unchanged when only pixel shuffle is used.

\begin{algorithm} \caption{First Encryption Scheme}
\begin{enumerate}
\item Read gray/color image.
\item Divide image into blocks with size $N \times N$.
\item Convert $2$D image block to $1$D block array by using Zigzag pattern.
\item Encrypt each $1$D block array with it's analogue secret key using XOR operation.
\item Repeat step $3$ and $4$ for each block and for each color component (R, G and B).
\end{enumerate}
\end{algorithm}

\section{The Proposed Image Encryption Schemes} \label{Sec3}
In this section, two encryption schemes for grayscale and color images are proposed. The first scheme is described in algorithm $1$ (Scheme $1$) and it starts with reading the grayscale or color image, then, divide it into blocks of size $N \times N$. Here, we divide the image into blocks of sizes $8 \times 8$, $16 \times 16$ and $32 \times 32$. Then, we convert each $2$D plain block to $1$D block array by using Zigzag pattern. The resulted block arrays are encrypted with it's analogue secret key using the XOR operation. We repeat the encryption operation until the end
of the plainimage blocks.

The second scheme starting the same way, but, the plainimage block is XORed with the previous encrypted image block before it is in turn encrypted according to the CBC mode given in Section \ref{Sec1c}. In the second scheme which described in algorithm $2$ (Scheme$2$), the encryption of each image block depends on all the previous blocks. In other words, each image block is used to modify the encryption of the next block. So, each cipherimage block is dependent not just on the plainimage block that generated it, but, on all the previous plainimage blocks.

\begin{algorithm} \caption{Second Encryption Scheme}
\begin{enumerate}
\item Read gray/color image.
\item Divide the image into blocks with size $N \times N$.
\item Convert $2$D image block to $1$D block array by using Zigzag pattern.
\item Encrypt the first $1$D block array with it's analogue secret key using XOR operation.
\item The resulting encrypted block from previous step is again XORed with the next plain block.
\item Repeat step $5$ for each next block and for each component.
\end{enumerate}
\end{algorithm}

{\begin{table}[h!]
\caption{Correlation coefficient for grayscale images.}
\label{tab-1}       
\begin{center}
\begin{tabular}{c| p{3em} | p{2.5em} | c | c | c} \cline{2-6}
& Block Size               & Gray Image                    & Horizontal & Vertical & Diagonal    \\ \hline
                                                
\multirow{9}{*}{Scheme$1$} & \multirow{3}{*}{$8 \times 8$}   & Lena1  & -0.110731 & 0.019723  & 0.064634  \\
                           &                                 & Brain1 & -0.078563 & 0.027462  & 0.068747  \\ 
                           &                                 & Brain2 & -0.080802 & 0.020420  & 0.059806  \\ \cline{2-6}
                           & \multirow{3}{*}{$16 \times 16$} & Lena1  & -0.003129 & 0.030368  & 0.054511  \\
                           &                                 & Brain1 & 0.001796  &	0.019956  & 0.047490  \\ 
                           &                                 & Brain2 & -0.001519 &	0.019827  &	0.044745  \\ \cline{2-6}
                           & \multirow{3}{*}{$32 \times 32$} & Lena1  & 0.015067  & -0.013705 & -0.008321 \\
                           &                                 & Brain1 & 0.006978  & -0.006833 & -0.013169 \\
                           &                                 & Brain2 & 0.009111  & -0.007917 & -0.005091 \\ \hline
\multirow{9}{*}{Scheme$2$} & \multirow{3}{*}{$8 \times 8$}   & Lena1  & -0.007554 &	-0.002917 & -0.002976 \\
                           &                                 & Brain1 & -0.008014 & 0.004825  &-0.001519  \\ 
                           &                                 & Brain2 & 0.015339  & 0.010465  & -0.006979 \\ \cline{2-6}
                           & \multirow{3}{*}{$16 \times 16$} & Lena1  & 0.001588  & 0.010809  &-0.004753  \\
                           &                                 & Brain1 & 0.002683  & 0.004188  & 0.000231  \\ 
                           &                                 & Brain2 & -0.001019 & 0.040991  & -0.009674 \\ \cline{2-6}
                           & \multirow{3}{*}{$32 \times 32$} & Lena1  & 0.008448  & 0.005313  & -0.002331 \\
                           &                                 & Brain1 & -0.000296 & 0.002431  & -0.002793 \\
                           &                                 & Brain2 & 0.002351  &-0.011532  &-0.001332  \\ \hline               
\end{tabular}
\end{center}
\end{table}} 

{\begin{table}[h!]
\caption{Correlation coefficient for color images.}
\label{tab-2}       
\begin{center}
\begin{tabular}{c| p{3em} | p{2.5em} | c | c | c} \cline{2-6}
& Block Size               & Color Image & Horizontal & Vertical  & Diagonal    \\ \hline
\multirow{9}{*}{Scheme$1$} & \multirow{3}{*}{$8 \times 8$}   & Lena2   & -0.082513 & 0.037293   & 0.068243    \\
                           &                                 & Lena    & -0.078532 & 0.036584   & 0.063751    \\ 
                           &                                 & Peppers & -0.049236 & 0.037812   & 0.031654  \\ \cline{2-6}
                           & \multirow{3}{*}{$16 \times 16$} & Lena2   & 0.010421  & -0.000204  & 0.045462  \\
                           &                                 & Lena    & 0.012032  & -0.000660  & 0.042805  \\ 
                           &                                 & Peppers & 0.011683  & -0.002831  & 0.031449  \\ \cline{2-6}
                           & \multirow{3}{*}{$32 \times 32$} & Lena2   & 0.019763  & -0.009177  & -0.012030 \\
                           &                                 & Lena    & 0.018391  & -0.011639  & -0.012637 \\
                           &                                 & Peppers & 0.011303  & -0.011811  &-0.004381  \\ \hline
\multirow{9}{*}{Scheme$2$} & \multirow{3}{*}{$8 \times 8$}   & Lena2   & -0.005078 & 0.009510   & 0.000686  \\
                           &                                 & Lena    & -0.011888 & -0.002805  & 0.005339  \\ 
                           &                                 & Peppers & 0.007140  & -0.000409  & 0.005576  \\ \cline{2-6}
                           & \multirow{3}{*}{$16 \times 16$} & Lena2   & -0.004184 & -0.002763  & 0.007892  \\
                           &                                 & Lena    & -0.003383 & 0.005404   & 0.007359  \\ 
                           &                                 & Peppers & 0.002886  & 0.001597   & 0.001081  \\ \cline{2-6}
                           & \multirow{3}{*}{$32 \times 32$} & Lena2   & 0.004221  & -0.004797  & 0.001803  \\
                           &                                 & Lena    & 0.005993  & -0.003371  & -0.005426 \\
                           &                                 & Peppers & 0.003712  & -0.010077  & -0.004915  \\ \hline               
\end{tabular}
\end{center}
\end{table}} 

\section{Experimental Results} \label{Sec4}
In this section, the performance of the two proposed schemes is analyzed by using different security test measures. These measures are taken as follows: key space analysis, statistical analysis including histogram analysis and computing the correlation coefficients of adjacent pixels, information entropy analysis, test security against differential attack including calculating the number of pixel change rate (NPCR) and unified average changing intensity (UACI). The used grayscale images are (brain1 and brain2) with size $256 \times 256$ and (lena1) with size $512 \times 512$. Also, the color images (lena and peppers) with size $256 \times 256$ and (lena2) with size $512 \times 512$ are used and the security analysis of the cipherimages is carried out.

\begin{figure*}
\centering
\includegraphics[width= 0.90\textwidth]{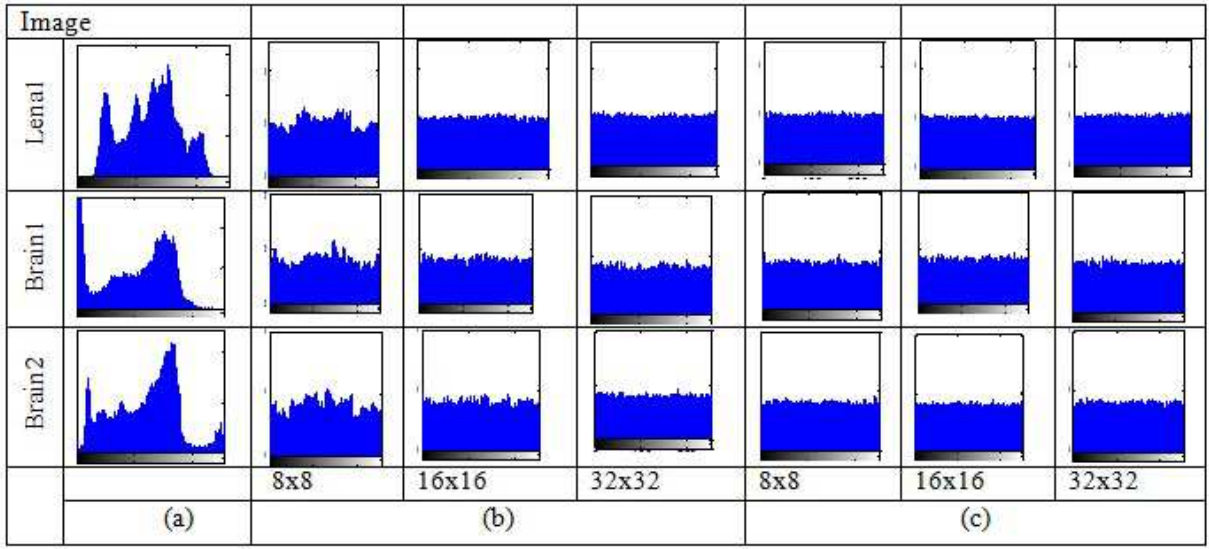}
\caption{Histogram of a) Original image, b) Scheme$1$ encrypted image, c) Scheme$2$ encrypted image, for different grayscale images.}
\label{fig-1}       
\end{figure*}

\subsection{Histogram Analysis} \label{Sec4.1}
\begin{figure*}
\centering
\includegraphics[width= 0.92\textwidth]{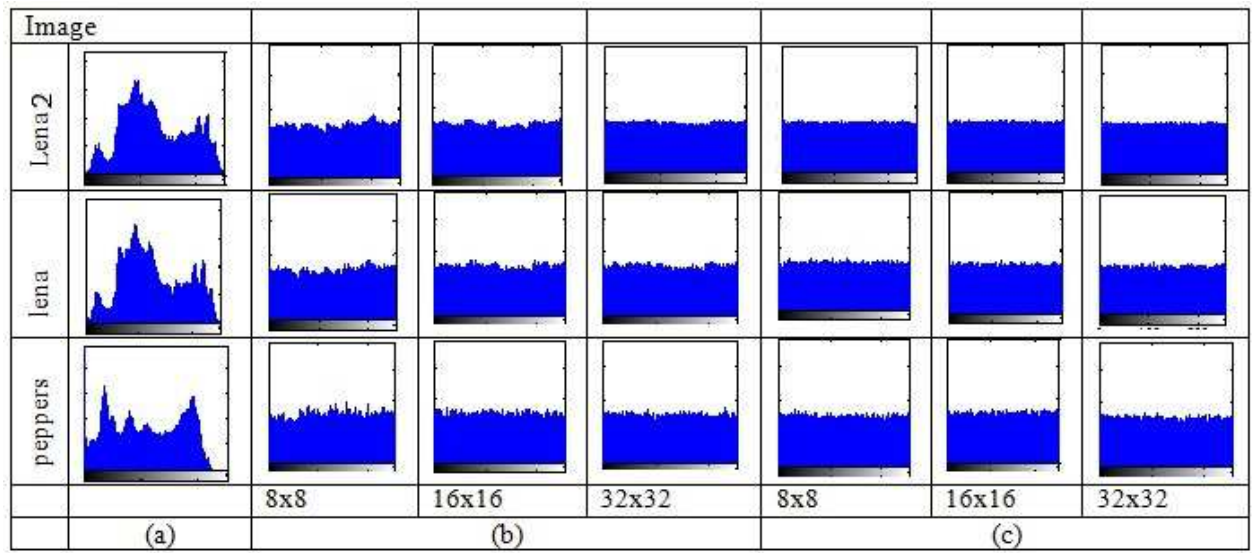}
\caption{Histogram of a) Original image, b) Scheme$1$ encrypted image, c) Scheme$2$ encrypted image, for different color images.}
\label{fig-2}       
\end{figure*}

To prevent the leakage of information, it is important to ensure that cipherimage does not have any statistical resemblance with it's original plainimage. A robust encryption scheme should always generate a cipherimage of uniform histogram for any plainimage. In this work, the histograms are plotted for grayscale/color plain and ciphered images. Figs. \ref{fig-1}(a -- c) and Figs. \ref{fig-2}(a -- c) display the histogram of the grayscale and color images (Fig. \ref{fig-1}(a) and Fig. \ref{fig-2}(a) and the corresponding cipherimages (Fig. \ref{fig-1}(b, c) and Fig. \ref{fig-2}(b, c)), respectively. From these figures, one can clearly notice that the histograms of the ciphered image are fairly uniform and significantly different about those of the original image. The statistical feature of the original images is enhanced in such a manner that the cipherimages had a uniform level distribution and good balance property.

\subsection{Correlation Analysis} \label{Sec4.2}
It is known that two adjacent pixels in every plainimage are highly correlated vertically, horizontally and diagonally. This could be the property of any ordinary image. The maximum value of correlation coefficient test is $1$ and the minimum value is $0$. A robust image encryption scheme versus statistical attack should have a correlation coefficient value of \~{}0. Results of horizontal, vertical and diagonal directions are obtained as shown in Table \ref{tab-1} for different grayscale images and Table \ref{tab-2} for different color images. These results demonstrate that there is negligible correlation between the two adjacent pixels in the cipherimages, even when these two adjacent pixels in the plainimage are highly correlated. Also, it is comparable to the correlation coefficient values presented by references \cite{c10},\cite{c14}, \cite{c18}, \cite{c19}, \cite{c20} and \cite{c21} as shown in Table \ref{tab-3}.

\subsection{Entropy Analysis} \label{Sec4.3}
\begin{figure*}
\centering
\includegraphics[width= 0.92\textwidth]{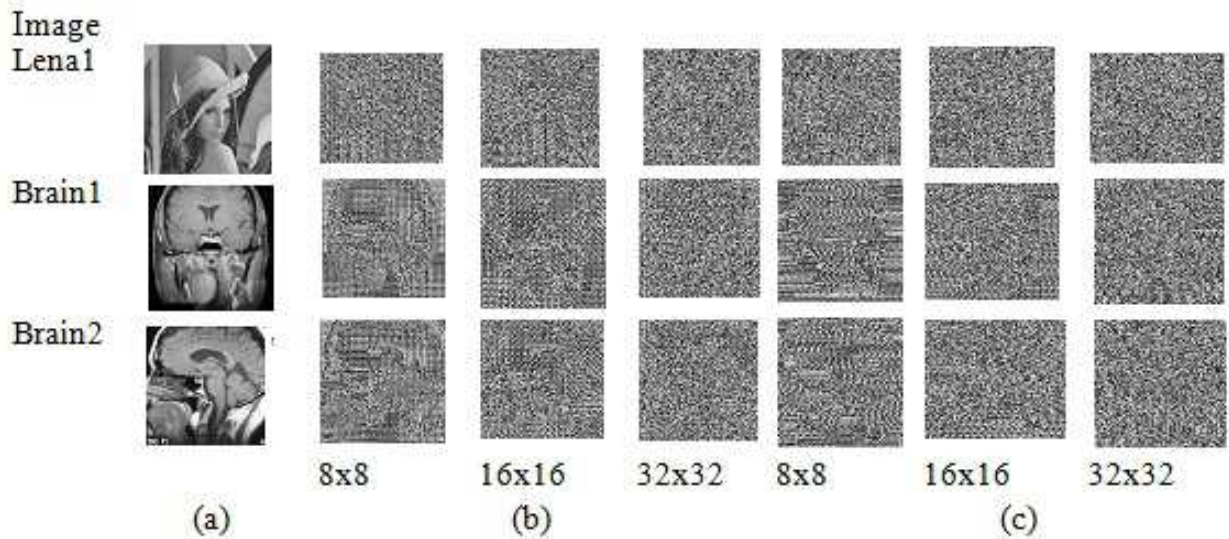}
\caption{Show the a) Original image, b) Scheme$1$ encryption image, c) Scheme$2$ encryption image, for different grayscale images.}
\label{fig-3}       
\end{figure*}

\begin{figure*}
\centering
\includegraphics[width= 0.92\textwidth]{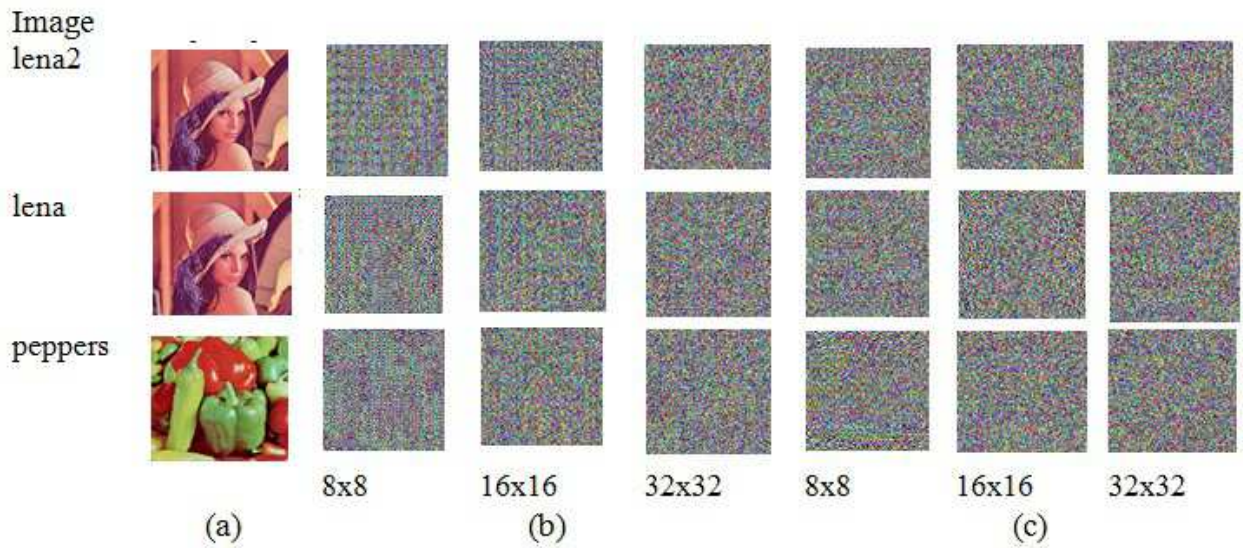}
\caption{Show the a) Original image, b) Scheme$1$ encryption image, c) Scheme$2$ encryption image, for different color images.}
\label{fig-4}       
\end{figure*}

Entropy is defined to show the degree of uncertainties in the system. It is well known that the entropy $H(m)$ of a message source $m$ can be calculated as:

\begin{equation}
\label{EQ9.1}
H{(m)} = - \sum\limits_{i=0}^{255} {P{(m_{i})}log_{2}}{P(m_{i})}
\end{equation}

\noindent where $P(m_{i})$ represents the probability of symbol $m_{i}$. For all the considered cipherimages shown in Fig. \ref{fig-3}(b, c) and Fig. \ref{fig-4}(b, c), the number of occurrence of each grayscale and color images is recorded and the probability of occurrence is computed for grayscale images and color images with different block sizes, respectively. Table \ref{tab-4} and \ref{tab-5} indicates the various values of the entropies for the encrypted images by the presented schemes. It can be noted that the entropy of the cipherimages are very near to the theoretical value of $8$ indicating that all the pixels in the encrypted images occur with almost equal probability. Therefore, the information leakage in the proposed encryption schemes is negligible, and it is secure against the entropy-based attack. Also it is comparable to the entropy values presented by references \cite{c12}, \cite{c14}, \cite{c18} and \cite{c22} as shown in Table \ref{tab-6}.

{\begin{table}[h!]
\caption{Comparison of Correlation Coefficient for Lena color image}
\label{tab-3}       
\begin{center}
\begin{tabular}{ l | c | c | c } \hline
Scheme                 & Horizontal & Vertical   & Diagonal    \\ \hline
Original Lena image    & 0.958853   & 0.980061   & 0.943422    \\ \hline
The proposed scheme$1$ & 0.017363   & -0.011263  & -0.012563    \\
The proposed scheme$2$ & -0.011888  & -0.002805  & 0.005331     \\
Ref.\cite{c10}         & 0.1257     & 0.0581     & 0.0504     \\
Ref.\cite{c14}         & 0.0681     & 0.0845     & 0.0046    \\
Ref.\cite{c18}         & -0.00124   & 0.00176    & 0.00193    \\
Ref.\cite{c19}         & -0.00368   & 0.00014    & -0.02298    \\
Ref.\cite{c20}         & 0.0042     & 0.0033     & 0.0024      \\ 
Ref.\cite{c21}         & -0.0018    & 0.00033    & 0.00427     \\ \hline
\end{tabular}
\end{center}
\end{table}} 

\subsection{Sensitivity Analysis} \label{Sec4.4}
In order to avoid the known-plaintext attack, the changes in the cipherimage should be significant even with a minor change in the plainimage. If one small change in the plainimage can cause a significant change in the cipherimage, with respect to diffusion and confusion properties, then the differential attack actually loses its efficiency and becomes practically useless. To quantify this requirement, two common measures are used here: number of pixels change rate (NPCR) and unified average changing intensity (UACI) \cite{c23}. We have tested the NPCR and UACI with the proposed encryption schemes to assess the influence of changing a single pixel in the plainimages on the cipherimages. From the obtained results, we have found that the average values of the percentage of pixels changed in cipherimages is greater than 99.65\% for NPCR and 33.46\% for UACI for the two proposed encryption schemes. This implies that the proposed schemes is very sensitive with respect to minor changes in the plainimage as shown in Table \ref{tab-7} and Table \ref{tab-8}. 

\subsection{Key Space Analysis} \label{Sec4.5}
The key space that is being used for encryption must be large enough to make the brute-force attack infeasible \cite{c24}. The key sequences generated using the elliptic curve generator over the field $\mathbb{F}_{2}$ had high periodicity so that the cipherimages are secure. In addition, the used chaos-driven elliptic curve pseudo-random key sequence generator has a flexible, moderately large key space, which comprises of the following parameters:

\begin{itemize}
\item[1:] The secret key of the chaotic generator (if the precision is $10^{-14}$, then, the size of the key space for
initial condition and control parameter is $2^{93}$), 
\item[2:] Possible elliptic curves and the base point,
\item[3:] The external secret user's key of CBC mode.
\end{itemize}

Then, the total number of possible keys is the size of the key space and is equal to the product of the above parameters. It is to be noted that unless all the above elements of the key space are known to the attacker, decryption using brute force attack is difficult. Even if the proposed schemes are hacked, after number of iterations and using different keys, the attacker is able to view only one single part/block of the image.

\section{Conclusion} \label{Sec5}
Image encryption algorithms play a vital role in the security of digital images and is considered one common method to protect the image information. In this paper, we presented two encryption schemes for grayscale/color images based on split the image into sub-blocks of different sizes $N \times N$ to increase the image security. Each block is transformed into a one dimensional ($1$D) array by using the Zigzag pattern. Then, the XOR logical operation is used to encrypt each block with the analogous secret key. In the second scheme, after the transformation process and before the next block is encrypted, it is XORed with the first encrypted block to become the next input to the encrypting routine and so on. This feedback mechanism depends on CBC mode of operation which considers highly nonlinear. 

The results show that lower correlation and higher entropy resulted by using smaller block sizes. Also, the results showed that the correlation between image elements was significantly decreased by using the second proposed scheme (Scheme$2$) with block size $16 \times 16$ and $32 \times 32$. 

{\begin{table}[h!]
\caption{Entropy values for grayscale images.}
\label{tab-4}       
\begin{center}
\begin{tabular}{l| l | c | c} \cline{2-4}
& Block Size               & Gray Image  & Entropy     \\ \hline
\multirow{9}{*}{Scheme$1$} & \multirow{3}{*}{$8 \times 8$}   & Lena1   & 7.987895      \\
                           &                                 & Brain1   & 7.986050       \\ 
                           &                                 & Brain2   & 7.983381      \\ \cline{2-4}
                           & \multirow{3}{*}{$16 \times 16$} & Lena1    & 7.99897       \\
                           &                                 & Brain1    & 7.99643        \\ 
                           &                                 & Brain2    & 7.996091       \\ \cline{2-4}
                           & \multirow{3}{*}{$32 \times 32$} & Lena1   & 7.999046       \\
                           &                                 & Brain1    & 7.99747        \\
                           &                                 & Brain2    & 7.998037        \\ \hline
\multirow{9}{*}{Scheme$2$} & \multirow{3}{*}{$8 \times 8$}   & Lena1   & 7.99914        \\
                           &                                 & Brain1    & 7.99747       \\ 
                           &                                 & Brain2    & 7.99782       \\ \cline{2-4}
                           & \multirow{3}{*}{$16 \times 16$} & Lena1    & 7.999268     \\
                           &                                 & Brain1    & 7.99768       \\ 
                           &                                 & Brain2    & 7.99842       \\ \cline{2-4}
                           & \multirow{3}{*}{$32 \times 32$} & Lena1   & 7.999294         \\
                           &                                 & Brain1    & 7.998186         \\
                           &                                 & Brain2    & 7.998090         \\ \hline               
\end{tabular}
\end{center}
\end{table}} 

{\begin{table}[h!]
\caption{Entropy values for color images.}
\label{tab-5}       
\begin{center}
\begin{tabular}{l| l | c | c} \cline{2-4}
& Block Size               & Color Image  & Entropy     \\ \hline
\multirow{9}{*}{Scheme$1$} & \multirow{3}{*}{$8 \times 8$}   & Lena2    & 7.9974      \\
                           &                                 & Lena     & 7.99653       \\ 
                           &                                 & Peppers  & 7.994405      \\ \cline{2-4}
                           & \multirow{3}{*}{$16 \times 16$} & Lena2    & 7.99854       \\
                           &                                 & Lena     & 7.99764       \\ 
                           &                                 & Peppers  & 7.99764       \\ \cline{2-4}
                           & \multirow{3}{*}{$32 \times 32$} & Lena2    & 7.99942       \\
                           &                                 & Lena     & 7.99869        \\
                           &                                 & Peppers  & 7.99871        \\ \hline
\multirow{9}{*}{Scheme$2$} & \multirow{3}{*}{$8 \times 8$}   & Lena2    & 7.99973        \\
                           &                                 & Lena     & 7.99903       \\ 
                           &                                 & Peppers  & 7.99873        \\ \cline{2-4}
                           & \multirow{3}{*}{$16 \times 16$} & Lena2    & 7.99978       \\
                           &                                 & Lena     & 7.99902       \\ 
                           &                                 & Peppers  & 7.99890       \\ \cline{2-4}
                           & \multirow{3}{*}{$32 \times 32$} & Lena2    & 7.99976       \\
                           &                                 & Lena     & 7.99897        \\
                           &                                 & Peppers  & 7.99876        \\ \hline               
\end{tabular}
\end{center}
\end{table}} 

{\begin{table}[h!]
\caption{Comparison of Entropy value for Lena color image.}
\label{tab-6}       
\begin{center}
\begin{tabular}{ l | c } \hline
Scheme                & Entropy    \\ \hline
The proposed scheme$1$  & 7.998137    \\
The proposed scheme$2$  & 7.999034    \\
Ref.\cite{c12}          & 7.985467    \\
Ref.\cite{c14}          & 7.975033    \\
Ref.\cite{c18}          & 7.989633    \\
Ref.\cite{c22}          & 7.9870      \\ \hline
\end{tabular}
\end{center}
\end{table}} 


{\begin{table}[h!]
\caption{NPCR and UACI of grayscale images}
\label{tab-7}       
\begin{center}
\begin{tabular}{l| l | c | c | c } \cline{2-5}
& Block Size               & Gray Image & NPCR (\%) & UACI (\%)      \\ \hline
\multirow{9}{*}{Scheme$1$} & \multirow{3}{*}{$8 \times 8$}   & Lena1  & 99.5712  & 27.3573 \\
                           &                                 & Brain1 & 99.5947  & 30.4378  \\ 
                           &                                 & Brain2 & 99.6005  & 28.8077  \\ \cline{2-5}
                           & \multirow{3}{*}{$16 \times 16$} & Lena1  & 99.6784  & 28.9233  \\
                           &                                 & Brain1 & 99.6503  & 32.1215  \\ 
                           &                                 & Brain2 & 99.6708  & 30.3371  \\ \cline{2-5}
                           & \multirow{3}{*}{$32 \times 32$} & Lena1  & 99.6093  & 33.4635  \\
                           &                                 & Brain1 & 99.5966  & 32.2349 \\
                           &                                 & Brain2 & 99.6113  & 30.6637 \\ \hline
\multirow{9}{*}{Scheme$2$} & \multirow{3}{*}{$8 \times 8$}   & Lena1  & 99.5853  & 28.6517 \\
                           &                                 & Brain1 & 99.6259  & 31.8237 \\ 
                           &                                 & Brain2 & 99.6015  & 30.2017  \\ \cline{2-5}
                           & \multirow{3}{*}{$16 \times 16$} & Lena1  & 99.6105  & 28.5656 \\
                           &                                 & Brain1 & 99.6230  & 31.9521 \\ 
                           &                                 & Brain2 & 99.5761  & 30.3085  \\ \cline{2-5}
                           & \multirow{3}{*}{$32 \times 32$} & Lena1  & 99.6124  & 28.6939 \\
                           &                                 & Brain1 & 99.6337  & 31.8058  \\
                           &                                 & Brain2 & 99.6425  & 30.2723 \\ \hline               
\end{tabular}
\end{center}
\end{table}} 

{\begin{table}[h!]
\caption{NPCR and UACI of color images }
\label{tab-8}       
\begin{center}
\begin{tabular}{l| l | c | c | c } \cline{2-5}
& Block Size               & Color Image & NPCR (\%) & UACI (\%)      \\ \hline
\multirow{9}{*}{Scheme$1$} & \multirow{3}{*}{$8 \times 8$}   & Lena2   & 99.6093   & 33.4635 \\
                           &                                 & Lena    & 99.7339   & 31.2896  \\ 
                           &                                 & Peppers & 99.7014   & 33.3449  \\ \cline{2-5}
                           & \multirow{3}{*}{$16 \times 16$} & Lena2   & 99.6093   & 33.4635 \\
                           &                                 & Lena    & 99.6515   & 31.0247 \\ 
                           &                                 & Peppers & 99.6383   & 32.2480 \\ \cline{2-5}
                           & \multirow{3}{*}{$32 \times 32$} & Lena2   & 99.6093   & 33.4635  \\
                           &                                 & Lena    & 99.6358   & 30.4976 \\
                           &                                 & Peppers & 99.5854   & 32.2896 \\ \hline
\multirow{9}{*}{Scheme$2$} & \multirow{3}{*}{$8 \times 8$}   &  Lena2  & 99.6093   & 33.4635 \\
                           &                                 & Lena    & 99.6114   & 30.3397 \\ 
                           &                                 & Peppers & 99.6164   & 32.1348  \\ \cline{2-5}
                           & \multirow{3}{*}{$16 \times 16$} & Lena2   & 99.6093   & 33.4635 \\
                           &                                 & Lena    & 99.5885   & 30.4506  \\ 
                           &                                 & Peppers & 99.6276   & 32.3855 \\ \cline{2-5}
                           & \multirow{3}{*}{$32 \times 32$} & Lena2   & 99.6093   & 33.4635 \\
                           &                                 & Lena    & 99.6007   & 30.4142 \\
                           &                                 & Peppers & 99.5905   & 32.0023 \\ \hline               
\end{tabular}
\end{center}
\end{table}} 


{\begin{table}[h!]
\caption{Comparison of NPCR and UACI for Lena color image.}
\label{tab-9}       
\begin{center}
\begin{tabular}{ l | c | c } \hline
Scheme                  & NPCR (\%) & UACI (\%)   \\ \hline
The proposed scheme$1$  & 99.6023   & 30.339    \\
The proposed scheme$2$  & 99.6514   & 33.463    \\
Ref.\cite{c10}          & 99.52     & 26.7933     \\
Ref.\cite{c14}          & 99.5843   & 33.3755    \\
Ref.\cite{c17}          & 99.6358   & 33.4428    \\
Ref.\cite{c18}          & 42.7519   & 13.2874    \\
Ref.\cite{c19}          & 99.7915   & 49.2191   \\
Ref.\cite{c20}          & 99.2173   & 33.4055     \\
Ref.\cite{c21}          & 99.9654   & 33.5720     \\
Ref.\cite{c25}          & 99.6062   & 33.8981    \\ \hline
\end{tabular}
\end{center}
\end{table}} 



%

\end{document}